\documentclass[aps,prd,
              reprint,
              amsmath,amssymb,amsfonts,groupedaddress,
              twocolumn,
              showpacs,
              superscriptaddress,floatfix]{revtex4-1}

\usepackage{graphicx}
\usepackage{dcolumn}
\usepackage{bm}

\usepackage{color}

\usepackage[breaklinks=true]{hyperref}
  \hypersetup{
  colorlinks=true,
  linkcolor=black,
  citecolor=black,
  filecolor=black,
  menucolor=black,
  urlcolor=black
  }
\usepackage{breakurl}

\begin{document}

\title{Electron dynamics controlled via self-interaction}

\author{Matteo Tamburini}\email{matteo.tamburini@mpi-hd.mpg.de}
\affiliation{Max-Planck-Institut f\"ur Kernphysik, Saupfercheckweg 1, D-69117 Heidelberg, Germany}
\author{Christoph H. Keitel}
\affiliation{Max-Planck-Institut f\"ur Kernphysik, Saupfercheckweg 1, D-69117 Heidelberg, Germany}
\author{Antonino Di Piazza}
\affiliation{Max-Planck-Institut f\"ur Kernphysik, Saupfercheckweg 1, D-69117 Heidelberg, Germany}

\date{\today}

\begin{abstract}
The dynamics of an electron in a strong laser field 
can be significantly altered by radiation reaction. 
This usually results in a strongly damped motion, with 
the electron losing a large fraction of its initial energy. 
Here we show that the electron dynamics in a bichromatic 
laser pulse can be indirectly controlled by a comparatively 
small radiation reaction force through its interplay with the Lorentz force. 
By changing the relative phase between the two frequency components 
of the bichromatic laser field, an ultrarelativistic electron bunch 
colliding head-on with the laser pulse can be deflected in a controlled way, 
with the deflection angle being independent of the initial electron energy. 
The effect is predicted to be observable with laser powers and intensities 
close to those of current state-of-the-art petawatt laser systems. 
\end{abstract}


\maketitle

\section{Introduction}

The rapid progress of high-power laser systems 
has paved the way for the investigation of unexplored regimes 
of laser-matter interaction with a number of applications, e.g., in 
extreme field physics~\cite{ehlotzkyRPP09,dipiazzaRMP12}, 
nuclear physics~\cite{ledinghamS03}, hadron-therapy~\cite{kraftNJP10,salaminPRL08} 
and relativistic laboratory astrophysics~\cite{bulanovEPJD09}. 
Next-generation 10-PW optical laser systems are expected to achieve 
intensities beyond $10^{23}\text{W/cm$^2$}$~\cite{dipiazzaRMP12,korzhimanovPU11}, 
and laser pulses with power beyond 100~PW and intensity up to 
$10^{25}\,\text{W/cm$^2$}$ are envisaged at the Extreme Light Infrastructure (ELI)~\cite{eliURL} 
and at the eXawatt Center for Extreme Light Studies (XCELS)~\cite{xcelsURL}. 
At such ultrahigh intensities, an electron becomes relativistic 
in a fraction of the laser period and its dynamics is 
dominated by radiation reaction (RR) effects, i.e., 
by the back reaction on the electron's motion 
of the radiation emitted by the electron itself while being accelerated 
by the laser pulse~\cite{landau-lifshitz-rr}. 
Hence, a deep understanding of RR effects is crucial 
for the design and the interpretation of future laser-matter experiments 
in the ultrarelativistic regime. 
Indeed, RR effects have several important implications ranging from 
the generation of high-energy photon~\cite{kogaPoP05,nakamuraPRL12,ridgersPRL12}, 
electron~\cite{michelPRE06,kostyukovPRSTAB12,keitelJPB98} and 
ion~\cite{naumovaPRL09,schlegelPoP09,tamburiniNJP10,chenPPCF11,tamburiniPRE12} 
beams, to the determination of bounds on particle acceleration in relativistic 
astrophysics~\cite{aharonianPRD02,medvedevPRE03}. 

At available and upcoming laser intensities, RR effects become large for ultrarelativistic electrons, 
where the RR force basically amounts to a strongly nonlinear 
and anisotropic friction-like force~\cite{tamburiniNJP10}. 
This explains why all the proposals to experimentally test 
the underlying equation of motion [the so called Landau-Lifshitz~(LL)~\cite{landau-lifshitz-rr} equation] 
rely on the RR-driven damping of the electron motion 
when an ultrarelativistic electron beam collides head-on 
with an intense laser pulse~\cite{kogaPoP05,dipiazzaPRL09,harveyPRD11a,harveyPRA12,thomasPRX12}. 
However, the research to date has focused on revealing RR effects 
and understanding their fundamental features rather than exploiting them 
in a possibly beneficial and controlled way. 

In this paper, we show that RR effects can provide 
a route to the control of the electron dynamics via 
the nonlinear interplay between the Lorentz and the RR force. 
This is achieved in a setup where an ultrarelativistic electron 
is exposed to a strong either few-cycle~\cite{tamburiniXXX12} 
or bichromatic~\cite{mironovQE11} laser pulse. 
Our exact analytical calculations for a plane-wave pulse and 
our more realistic numerical simulations for a focused laser pulse 
show that, already at the intensities achievable with state-of-the-art laser systems, 
an ultrarelativistic electron colliding head-on with a bichromatic laser pulse 
can be deflected in an ultrafast and controlled way within a cone 
of about $8^\circ$ aperture \emph{independently} of the initial electron energy 
as long as quantum effects remain small. 
At still higher intensities, the interplay between the RR 
and the Lorentz force can even overcome the radiation losses themselves, 
resulting in a RR assisted electron acceleration instead of damping.

\section{Electron dynamics in an arbitrary plane-wave field}

The LL equation of an electron (mass $m$ and charge $e$) in the presence
of an external electromagnetic field $F^{\mu\nu}$,
is~\cite{landau-lifshitz-rr}:
\begin{equation} \label{eq1}
\frac{d u^{\mu}}{d \tau} = - F^{\mu\nu} u_{\nu} + r_{R} \left[ 
F^{\mu\nu} F_{\nu\alpha} u^{\alpha} - (F^{\beta\nu} u_{\beta} F_{\nu\alpha} u^{\alpha}) u^{\mu}
\right],
\end{equation}
where $\tau$ is the proper time, $u^{\mu}\equiv dx^{\mu}/d\tau$ and 
where $r_{R}=4\pi e^2/3mc^2\lambda\approx1.18\times10^{-8}/\lambda_{\text{$\mu$m}}$,
with $\lambda$ being a typical length scale, conveniently
chosen as the wavelength of a Ti:sapphire laser, i.e., $\lambda=0.8\,\text{$\mu$m}$. 
In Eq.~(\ref{eq1}) dimensionless units have been employed, such that
time is in units of $\omega^{-1}\equiv \lambda/2\pi c$, 
length is in units of $\omega^{-1}c$, and fields are in units of $E^*\equiv m \omega c/|e|$. 
Note that the term of the RR force containing the derivatives of the field tensor $F^{\mu\nu}$~\cite{landau-lifshitz-rr} 
has been neglected in Eq.~(\ref{eq1}) since its contribution is smaller than quantum effects~\cite{tamburiniNJP10} 
and it does not appreciably influence the electron dynamics in the regime of interest here. 

Modeling the laser pulse as a plane wave propagating along 
the direction $\vec{n}$, the LL equation can be solved exactly for 
any plane-wave electromagnetic field which is an \emph{arbitrary} function 
of the phase of the wave $\varphi=n_\mu x^\mu$ only, where 
$n^\mu\equiv (1,\vec{n})$ and $n^\mu n_\mu=0$~\cite{dipiazzaLMP08}. 
Hereafter, the subscripts $0$ and $f$ refer to the initial and final value 
of the corresponding quantity, respectively. 
In order to analyze the origin of each term in the solution, 
we first omit the last term on the right-hand side of Eq.~(\ref{eq1}) (Larmor term).  
In this case $d\tau/d\varphi=1/\rho_0$, where $\rho_0\equiv n_\mu u_0^\mu$ is 
the initial Doppler factor and $u_0^\mu$ is the initial four-velocity. 
Inclusion of the Larmor term renders the relation between the proper time 
and the phase nonlinear~\cite{dipiazzaLMP08}: $d\tau/d\varphi=h(\varphi)/\rho_0$ 
where $h(\varphi)=1+r_{R}\rho_0\int_{\varphi_0}^{\varphi}{[\vec{E}(\phi) 
\times\vec{B}(\phi)]\cdot\vec{n}\,d\phi}$, with 
$\vec{E}(\phi)$ and $\vec{B}(\phi)$ being the plane 
wave electric and magnetic field, respectively. 
For an arbitrary plane wave, Eq.~(\ref{eq1}) 
written as a function of the phase $\varphi$ becomes: 
\begin{equation} \label{eq2}
\frac{d\tilde{u}^\mu}{d\varphi}=-\frac{h}{\rho_0} F^{\mu\nu}\tilde{u}_\nu 
+ \frac{h}{\rho_0^2} \frac{d h}{d\varphi} n^\mu,
\end{equation}
where $\tilde{u}^\mu\equiv d x^\mu/d\varphi$. 
In Eq.~(\ref{eq2}), the only effect of the Larmor term is to multiply 
the terms in the right-hand side by $h(\varphi)$. 
Since $n_\mu F^{\mu\nu}=0$, the solution of Eq.~(\ref{eq2}) is the sum of 
the two solutions obtained considering each term 
on the right-hand side of Eq.~(\ref{eq2}) separately. 
The second term in Eq.~(\ref{eq2}) results in a contribution proportional 
to $(h^2-1)n^\mu$. This term accounts for the effect 
of the radiation pressure~\cite{landau-lifshitz-rr} 
and leads to a small energy gain when an electron at rest is swept 
by a laser pulse~\cite{fradkinPRL79,lehmannPRE11}. 
Finally, the first term on the right-hand side of Eq.~(\ref{eq2}) is 
a strongly nonlinear effective Lorentz force. 
This term can be integrated analytically, and 
the exact solution of Eq.~(\ref{eq2}) for 
the dimensionless four-momentum $p^{\mu}=(\varepsilon,\vec{p})$ 
as a function of $\varphi$ is~\cite{dipiazzaLMP08}:
\begin{eqnarray}
\varepsilon&=&\frac{\varepsilon_0}{h}+\frac{2\vec{\mathcal{I}}\cdot\vec{p}_0+
(h^2-1)+\vec{\mathcal{I}}^2}{2\rho_0 h}, \label{eq3} \\
\vec{p}&=&\frac{\vec{p}_0+\vec{\mathcal{I}}}{h}+
\frac{2\vec{\mathcal{I}}\cdot\vec{p}_0+
(h^2-1)+\vec{\mathcal{I}}^2}{2\rho_0 h}\vec{n}, \label{eq4}
\end{eqnarray}
where $\vec{\mathcal{I}}(\varphi)=-\int_{\varphi_0}^{\varphi}{h(\phi)\vec{E}(\phi)d\phi}$. 
Since $\vec{E}\cdot\vec{n}=0$, 
in Eq.~(\ref{eq4}) the vectors directed along $\vec{n}$ and $\vec{\mathcal{I}}$ 
correspond to the longitudinal and transverse momentum gain, respectively.

Let us consider a bichromatic plane-wave pulse propagating along 
the positive $z$-axis and polarized along the $x$-axis with 
$E_x(\varphi)=g(\varphi)\left[\xi_1\sin(\varphi+\theta_1)+\xi_2\sin(2\varphi+\theta_2)\right]$, 
where $g(\varphi)$ is a smooth temporal envelope identically vanishing 
for $\varphi$ outside the interval $(\varphi_0,\varphi_f)$, $\xi_1$, $\xi_2$ are the field amplitudes 
of each frequency component, and $\theta_1$, $\theta_2$ are two constant initial phases.
After the electron passes through the laser beam, the relevant functions
in the electron four-momentum are: 
$h_f=1+r_R\rho_0\Psi$, $\mathcal{I}_{y,f}=0$ and $\mathcal{I}_{x,f}=-r_R\rho_0\Delta$,
where 
$\Psi\equiv\int_{\varphi_0}^{\varphi_f}{d\phi \, E_x^2(\phi)}$, 
$\Delta\equiv\int_{\varphi_0}^{\varphi_f}{d\phi \, E_x(\phi) \int_{\varphi_0}^{\phi}{d\vartheta \, E_x^2(\vartheta)}}$.  
For simplicity, in the following we assume a pulse envelope 
$g(\varphi)=\sin^2(\varphi/2N)$ in the interval $(0,\varphi_f)$ (i.e. $\varphi_0=0$), 
where $N=\varphi_f/2\pi$ is the total integer number of cycles of the pulse. 

For the sake of comparison, we first consider a quasi-monochromatic plane wave ($\xi_2=0$ and $N\gg 1$). 
In this case, two frequencies are basically present in $E_x^2(\varphi)$, 
which arise from $\sin^2(\varphi+\theta_1)$. After integrating $E_x^2(\varphi)$, 
only the zero-frequency component provides a net contribution to $\Psi$. 
Analogously, the integrand of $\Delta$ only contains frequencies which are 
odd multiples of the central frequency $\omega$, and $\Delta$ averages out to zero for a quasi-monochromatic plane wave. 
In fact, in our case 
\begin{equation} \label{eq5}
\Delta = \frac{3\pi \xi_{1}^{3}N\cos(\theta_{1})}{16(N^2-1)} \qquad
\left(N\geq 4\right),
\end{equation}
which tends to zero for $N\to\infty$. 
The situation is essentially different for the bichromatic plane wave considered above. 
Here, a zero-frequency term arises in the integrand of $\Delta$, 
such that $\Delta$ \emph{diverges} in the limit $N\gg 1$:
\begin{equation} \label{eq6}
\Delta \approx \frac{15\pi}{64} \xi_{1}^{2}\xi_{2}N\cos(\theta_2-2\theta_1)
\qquad
\left(N\gg 1\right). 
\end{equation}
Recalling that $\mathcal{I}_{x,f}=-r_R\rho_0\Delta$, Eqs.~(\ref{eq4}) and~\ref{eq6}) 
already show \emph{in general} that the electron dynamics 
can be controlled either by changing the constant initial phase $(\theta_2-2\theta_1)$ 
or the field amplitudes $\xi_1,\,\xi_2$, and the effect 
dramatically increases for increasing $\xi_1,\,\xi_2,\,N$. 
Indeed, for $N\gg 1$ a different pulse envelope $g(\varphi)$ only alters 
the numerical factor on the right side of Eq.~(\ref{eq6}).  
Finally, we mention that $\mathcal{I}_{x,f}$ can become large also for 
ultraintense nearly one-cycle laser pulses~\cite{tamburiniXXX12}. 
However, in this case $\mathcal{I}_{x,f}$ is sensitive both to the 
carrier envelope phase $\theta_1$ and to the precise shape of the pulse $g(\varphi)$. 

Physically, without RR the electron transverse momentum 
$\vec{p}_{\perp}(\varphi)=\vec{p}(\varphi)-[\vec{n}\cdot\vec{p}(\varphi)]\vec{n}$ 
oscillates with the same frequencies as the plane-wave field [see Eq.~(\ref{eq4}) with $h(\varphi)=1$]. 
Hence, the cumulative effect of the force eventually averages out to zero. 
However, the energy loss associated with the RR force modulates the position 
of the electron within the plane-wave field.
For a quasi-monochromatic plane wave, there is no control on this 
modulation and thus no net transverse momentum gain~\cite{dipiazzaPRL09,thomasPRX12}, as the modulation 
is intrinsically related to the frequency of the driving field. 
On the contrary, if a higher-frequency field is also included, its frequency and absolute 
phase can be chosen in such a way that a Fourier component is nonlinearly generated in the 
resulting modulation, which resonantly oscillates with the lower-frequency field. 
In turn, this resonance can result in a net transverse momentum gain 
$\delta p_x=\mathcal{I}_{x,f}/h_f$, and the interplay of the two components 
of the bichromatic field is indeed reflected in Eq.~(\ref{eq6}).

\section{Electron dynamics control}

Let us consider the effects arising from the interaction 
of an ultrarelativistic electron colliding head-on with a second-harmonic 
enriched laser pulse. Hereafter, the term $(h^2-1)$ in the numerator 
of Eqs.~(\ref{eq3}) and (\ref{eq4}) is neglected in the analytical results, 
since it does not appreciably affect our conclusions. 
Eq.~(\ref{eq4}) indicates that the initially counterpropagating electron 
is deflected in the $xz$-plane asymmetrically. 
Since $\rho_0\approx 2|\vec{p}_0|$, the deflection angle  
with respect to the initial propagation direction is 
\begin{equation} \label{eq7}
\zeta\approx -\arctan\left(\frac{2r_R\Delta}{1-r_R^2\Delta^2}\right)
\end{equation}
if $r_R|\Delta|<1$, $\zeta+\pi$ if $r_R\Delta<-1$ and $\zeta-\pi$ if $r_R\Delta>1$ 
\emph{independently} of the initial electron energy. 
Again in the ultrarelativistic regime, for $r_R |\Delta|>1$ 
the electron is back reflected by the plane-wave pulse. 
We stress that this condition is independent of the initial electron energy 
because higher initial energies imply higher RR effects, 
the functions $h_f$ and $\vec{\mathcal{I}}_f$ being proportional 
to the initial Doppler factor $\rho_0$. 
In other words, for $r_R|\Delta|>1$ the laser pulse behaves like 
a perfectly reflecting electron ``mirror'', i.e., it reflects back 
all the electrons with arbitrarily high initial energy, 
as long as the onset of quantum effects does not severely alter the 
predictions of classical electrodynamics (see below). 
In addition, from Eq.~(\ref{eq3}) it follows that if the initial electron 
energy $\varepsilon_0$ is less than $r_R\Delta^2/2\Psi$ then a surprising
circumstance occurs: the final electron energy is 
\emph{larger} than its initial energy. 
In fact, although the direct effect of the RR force is to reduce the electron energy, 
it also alters the temporal electron evolution, such that the electron's world line 
with RR effects differs from the electron's world line without them. 
As a result, while without RR effects the Lorentz force cannot perform a net work on the electron
[see Eq.~(\ref{eq3}) with $h(\varphi)=1$], with RR effects the Lorentz force can perform a positive work 
along the \emph{RR-altered} electron world line. Hence, the dissipative RR force indirectly allows
the Lorentz force to accelerate the electron, and when $\varepsilon_0<r_R\Delta^2/2\Psi$ 
the indirect energy gain is larger than the direct energy loss. 
In order to observe this effect, an intensity beyond $10^{23}\;\text{W/cm$^2$}$ 
and a waist radius of the order of some tens of micrometers are required, resulting 
in a power of the order of a few exawatts. Although such powers are well beyond 
those currently available, they may be achieved employing 
coherent beam superposition techniques~\cite{xcelsURL,bagayevAIP12,mourouNP13}.

\section{Numerical results for a focused laser pulse}

The above analytical predictions are exact if the laser field is modeled as a plane wave. 
In order to test them in a more realistic set-up, we solve Eq.~(\ref{eq1}) 
numerically for a focused laser pulse interacting with an electron bunch. 
Our simulations show that the plane-wave and the focused pulse results
are in good agreement already with a $5\mu\text{m}$ waist radius (see below).
Following Refs.~\cite{mcdonaldXXX97,salaminPRL02}, 
a hyperbolic secant temporal envelope and a Gaussian transverse profile 
with terms up to the fifth order in the diffraction angle are employed 
to accurately describe the laser pulse, which reaches its maximal focusing 
at the origin with waist radius $w_{O}$. 
According to the notation employed so far, the laser beam stems from
two pulses with wavelengths $0.8\;\mu\text{m}$ and $0.4\;\mu\text{m}$,
respectively, and with peak field amplitudes $\xi_1$ and $\xi_2$, respectively. 
Hereafter, for simplicity we set the constant phase $\theta_1=0$.
The electrons are initially distributed according to a six-dimensional Gaussian 
probability distribution 
\begin{figure}
\begin{center}
\includegraphics[width=0.5\textwidth]{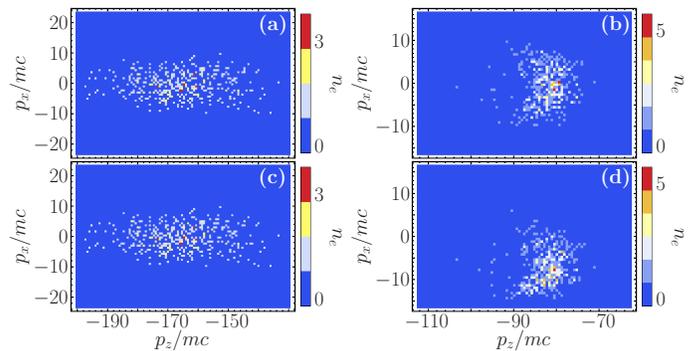}
\caption{(Color online) Electron density distribution $n_e(p_z,p_x)$ as a function of the longitudinal $p_z$ 
and transverse $p_x$ momentum after the interaction of 400~electrons 
with a bichromatic laser pulse. 
Panel~(a): $\cos(\theta_2)=0$ without RR. 
Panel~(b): $\cos(\theta_2)=0$ with RR. 
Panel~(c): $\cos(\theta_2)=1$ without RR. 
Panel~(d): $\cos(\theta_2)=1$ with RR. 
See the text for further numerical details.} \label{Fig1}
\end{center}
\end{figure}
\begin{equation}
f(\vec{x},\vec{p})=N_e\frac{e^{-\big[\frac{x^2+y^2}{2\sigma_{T}^2}+\frac{(z-z_0)^2}{2\sigma_{L}^2}\big]
-\big[\frac{p_{x}^2+p_{y}^2}{2\sigma_{p_T}^2}+\frac{(p_z-p_{z,0})^2}{2\sigma_{p_L}^2}\big]}}
{(2\pi)^3\sigma_T^2\sigma_{p_T}^2\sigma_L\sigma_{p_L}},
\end{equation}
with $N_e$ being the total number of electrons and 
$\sigma_T$ and $\sigma_L$ ($\sigma_{p_T}$ and $\sigma_{p_L}$) being the 
transverse and the longitudinal position (momentum) widths, respectively.

\subsection{Simulation setup}

In our simulation, the laser pulse is 70~fs long between its first 
and last half maximal intensity with $\xi_1=40$ ($3.4\times10^{21}\;\text{W/cm$^2$}$), 
$\xi_2=28$ ($1.7\times10^{21}\;\text{W/cm$^2$}$) and the waist radius is $w_{O}=5\;\mu\text{m}$. 
Hence, the total intensity and power are $5.1\times10^{21}\;\text{W/cm$^2$}$ 
and 2~PW, respectively. 
Initially, the electron bunch has mean momentum $p_{z,0}=-165\;mc$ with standard deviations 
$\sigma_{T}=0.2\;\mu\text{m}$, $\sigma_{L}=0.5\;\mu\text{m}$, $\sigma_{p_T}=1\;mc$ 
and $\sigma_{p_L}=12\;mc$. The electron average density is $3\times10^{15}\text{cm$^{-3}$}$ 
so that the electron bunch contains about 400~electrons. 
The above-mentioned laser parameters are similar to those 
of available petawatt laser systems~\cite{dipiazzaRMP12,korzhimanovPU11}. 
Much larger effects can be achieved at higher intensities, 
since the transverse momentum gain increases rapidly with rising 
laser field amplitudes $\xi_1,\,\xi_2$ [see Eq.~(\ref{eq6})]. 
In addition, the electron deflection can be controlled by changing 
either the phase $\theta_2$ or the amplitudes $\xi_1,\,\xi_2$. 
The latter approach can be exploited tuning the ratio 
between $\xi_1$ and $\xi_2$ by controlling the second-harmonic 
conversion efficiency, e.g., by changing the tilt angle 
in a tilted-crystal configuration~\cite{moriJAP98}.
To date, frequency-doubling efficiencies up to 73\% at 2~TW/cm$^2$ intensity
have been demonstrated experimentally for femtosecond pulses~\cite{mironovQE11}. 
Also, phase-control of bichromatic laser pulses has been employed 
at intensities of the order of $10^{14}\;\text{W/cm$^2$}$ to steer 
the electron dynamics in nonrelativistic atomic physics~\cite{watanabePRL94,*mauritssonPRL06,*xuAPB11}. 
Similar techniques might be extended to higher intensities 
via coherent beam superposition of multiple laser beams~\cite{xcelsURL,bagayevAIP12,mourouNP13}, 
since a relatively compact optics can be employed for each amplification channel. 
Finally, electron bunches with the same parameters as in our simulation 
have been generated experimentally employing standard multiterawatt optical lasers~\cite{lundhNP11}. 
Such relatively low-power pulses can also be generated 
by extracting a fraction of energy from the initial strong pulse
before the frequency-doubling.

\subsection{Results and discussion}

Figure~\ref{Fig1} reports the electron density distribution $n_e(p_z,p_x)$ 
as a function of the longitudinal $p_z$ and transverse $p_x$ momentum for 
the interaction of 400~electrons with the focused laser pulse both for 
$\cos(\theta_2)=0$ and $\cos(\theta_2)=1$, with and without RR. 
No appreciable difference between $\cos(\theta_2)=0$ and $\cos(\theta_2)=1$ 
is found if only the Lorentz force is taken into account. Furthermore, 
if the RR force is neglected, the mean of the momentum distribution 
remains unaltered after the electron bunch has passed through 
the laser pulse $\bar{p}_x\approx0$ and $\bar{p}_{z}\approx-165\;mc$ 
[see Figs.~\ref{Fig1}(a),~\ref{Fig1}(c)]. 
However, if the RR force is taken into account, 
for $\cos(\theta_2)=0$ the electrons still move along 
their initial propagation direction and are distributed symmetrically 
in the transverse momentum space with $\bar{p}_x\approx0$ and 
$\bar{p}_{z}\approx-82\;mc$ [see Fig.~\ref{Fig1}(b)] 
in good agreement with the plane wave prediction 
$p_{x,f}\approx0$ and $p_{z,f}\approx-79\;mc$. 
On the other hand, for $\cos(\theta_2)=1$ all the electrons are 
deflected in the transverse direction independently of their initial energy, 
the mean of the momentum distributions being $\bar{p}_x\approx-7\;mc$ 
and $\bar{p}_{z}\approx-82\;mc$ [see Fig.~\ref{Fig1}(d)]. 
For the corresponding plane-wave pulse, we obtain 
$p_{x,f}\approx-5.8\;mc$ and $p_{z,f}\approx-79\;mc$, 
in good agreement with the above mentioned focused pulse results. 

The effect of QED corrections to the classical prediction
has been estimated by introducing a quantum corrected RR force, 
which accounts for the reduction of the emitted power in the quantum case 
compared to the classical one~\cite{Baier-book}. 
The present approach is valid as long as the quantum parameter 
$\chi=|e|\hbar\sqrt{|[F^{\mu\nu}p_\nu]^2|}/m^3c^4$ (Gaussian units) 
remains much smaller than unity~\cite{dipiazzaRMP12,Baier-book}. 
Indeed, in our simulations we found $\chi\lesssim0.04$. 
Moreover, due to RR effects, $\chi$ remains significantly smaller 
compared to the case without RR, especially at higher laser pulse intensities. 
In our simulation, quantum corrections do not qualitatively affect the results 
but induce a correction to the final mean momenta of the electron distribution, 
with $\bar{p}_{x}\approx0$ and $\bar{p}_{z}\approx-87\;mc$ for $\cos(\theta_2)=0$, 
and $\bar{p}_x \approx-6\;mc$ and $\bar{p}_{z}\approx-88\;mc$ for $\cos(\theta_2)=1$. 
Finally, stochasticity effects in quantum RR may broaden the 
final electron distribution but do not significantly alter its mean value~\cite{neitzPRL13}. 
\begin{figure}
\begin{center}
\includegraphics[width=0.48\textwidth]{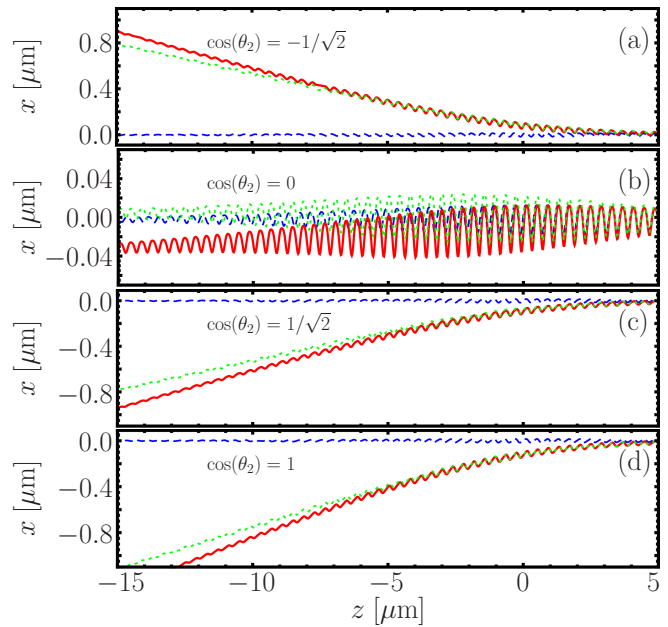}
\caption{(Color online) Trajectory of an electron colliding head-on with 
a bichromatic laser pulse without (blue dashed line) and 
with (red solid line) RR force included. The corresponding plane wave result 
with RR (green dotted line) is also reported for comparison. 
In all cases $\theta_1=0$. 
Panel~(a): $\cos(\theta_2)=-1/\sqrt{2}$. 
Panel~(b): $\cos(\theta_2)=0$. 
Panel~(c): $\cos(\theta_2)=1/\sqrt{2}$. 
Panel~(d): $\cos(\theta_2)=1$. 
See the text for further numerical details.} \label{Fig2}
\end{center}
\end{figure}

Figure~\ref{Fig2} displays the trajectory of an electron injected into the focus 
of the bichromatic laser pulse with initial momentum $\vec{p}_0=(0,0,-165\;mc)$
without (blue dashed line) and with (red solid line) RR effects included 
[the corresponding plane wave result with RR (green dotted line) is also shown for comparison]. 
In all cases, the electron passes through the laser pulse without changing its 
initial propagation direction when RR effects are neglected. 
When RR effects are included, for $\cos(\theta_2)=0$  
the electron goes through the laser pulse without 
significantly deviating from its initial propagation direction [see Fig.~\ref{Fig2}(b)], 
whereas it is quickly deflected in the transverse direction for $\cos(\theta_2)\neq0$ 
[see Figs.~\ref{Fig2}(a), \ref{Fig2}(c) and \ref{Fig2}(d)]. 
From Eq.~(\ref{eq7}) with $\cos(\theta_2)=1$ [$\cos(\theta_2)=\mp1/\sqrt{2}$], 
the predicted deflection angle for the plane-wave becomes 
$\zeta\approx-4.2^\circ$ ($\zeta\approx\pm3^\circ$) in fair agreement 
with the focused pulse result $\zeta\approx-5.4^\circ$ ($\zeta\approx\pm3.8^\circ$). 
Quantum effects lead to relatively small corrections, 
the deflection angle being $\zeta\approx-3.6^\circ$ ($\zeta\approx\pm2.5^\circ$) 
for the plane wave with $\cos(\theta_2)=1$ [$\cos(\theta_2)=\mp1/\sqrt{2}$] and 
$\zeta\approx-4.5^\circ$ ($\zeta\approx\pm3.2^\circ$) for the focused pulse. 

\begin{acknowledgments}
We acknowledge useful discussions with N.~Neitz, G.~Sarri, and A.~M.~Sergeev.
\end{acknowledgments}

\bibliography{Interplay}

\end{document}